# Band-asymmetry-driven nonreciprocal electronic transport in a helimagnetic semimetal α-EuP$_3$

Alex Hiro Mayo[1]*, Darius-Alexandru Deaconu[2], Hidetoshi Masuda[1], Yoichi Nii[1], Hidefumi Takahashi[3], Rodion Vladimirovich Belosludov[1], Shintaro Ishiwata[3], Mohammad Saeed Bahramy[2], and Yoshinori Onose[1]

[1]Institute for Materials Research, Tohoku University; Aoba-ku, Sendai, 980-8577, Japan.

[2]Department of Physics and Astronomy, School of Natural Sciences, The University of Manchester, Oxford Road, Manchester M13 9PL, United Kingdom

[3]Division of Materials Physics, Graduate School of Engineering Science, Osaka University, Toyonaka, Osaka 560-8531, Japan.

*Corresponding author. Email: alex.hiro.mayo.d1@tohoku.ac.jp

**Abstract**

Chiral magnetic textures give rise to unconventional magnetotransport phenomena such as the topological Hall effect and nonreciprocal electronic transport. While the correspondence between real-space magnetic topology/symmetry and such transport phenomena has been well established, a microscopic understanding based on the spin-dependent band structure in momentum space remains elusive. Here we demonstrate how a chiral magnetic structure in real space introduces an asymmetry in the electronic band structure and triggers a nonreciprocal electronic transport in a centrosymmetric helimagnet α-EuP$_3$. The magnetic structure of α-EuP$_3$ is highly tunable by a magnetic field and closely coupled to its semi-metallic electronic band structure, enabling a systematic study across chiral and achiral magnetic phases on the correspondence between nonreciprocal transport and electronic band asymmetry. Our findings reveal how a microscopic change in the magnetic configuration of charge carriers can lead to nonreciprocal electronic transport, paving the way for designing chiral magnets with desirable properties.

**Introduction**

Chirality is the breaking of mirror symmetry. The concept of chirality is applicable in a wide range of scientific fields, from high-energy physics to biology(1). In condensed matter, the chirality caused by the crystal structure is primarily studied. However, there is a distinct class of materials where the magnetic structure induces chirality. In particular, in helimagnets, the spiral arrangement of magnetic moments breaks the mirror symmetry, thus inducing chirality. The topological magnetic textures denoted as a skyrmion lattice can be viewed as the superposition of several helimagnetic structures(2, 3). The chiral magnetic textures give rise to a number of unique physical properties, among which the nonreciprocal electronic transport has been extensively studied recently(4–7).

Nonreciprocal electronic transport is the rectification of electric conduction in systems without spatial-inversion ($\mathscr{P}$) and time-reversal ($\mathscr{T}$) symmetries(8–12). When the crystal structure breaks $\mathscr{P}$, the mechanism is well-understood. For example, in BiTeBr(13), the polar crystal structure induces a Rashba-type spin splitting in the electronic structure. Then, an applied magnetic field gives rise to the asymmetry of the electronic band structure, causing a rectification effect. On the other hand, in chiral magnetic textures with a centrosymmetric crystal structure, the arrangement of the spins breaks $\mathscr{P}$ in a distinct manner such that it gives rise to a finite chirality in the system. The nonreciprocal electronic transport has been studied in chiral magnets such as MnP(6), MnAu$_2$(7), showing a sizeable effect of magnetic chirality on electronic transport. However, the mechanism of nonreciprocal electronic transport in the helimagnetic systems remains unclear.

In this article, through a combination of nonreciprocal electronic transport experiment and first-principles band calculation, we demonstrate that the nonreciprocal electronic transport is induced by the asymmetry of the band structure in a helimagnetic semimetal α-EuP$_3$. α-EuP$_3$ has a centrosymmetric crystal structure with a monoclinic space-group type *C*2/*m*(14), characterized by a single mirror plane perpendicular to the twofold rotational *b*-axis (Fig. 1A). The 4*f*-states of the divalent Eu ions are highly localized and relatively close in energy to the Fermi level, thus behave as local magnetic moments interacting through carrier-mediated Ruderman-Kittel-Kasuya-Yosida coupling(15–17).

**Results and Discussion**

It has been recently noted that α-EuP$_3$ exhibits a characteristic interplay between magnetism and electronic structure, leading to unusual magnetotransport responses such as a large anomalous Hall effect caused by a topological transition(14). However, the discussion was mainly for the paramagnetic regime, and therefore the transport involving the magnetically ordered regime remains unraveled. Figure 1B shows the field-temperature phase diagram of α-EuP$_3$ based on resistivity and magnetization measurements under magnetic fields applied along the *a*-axis. Two low-temperature magnetic phases are observed, consistent with the previously reported results of isostructural P-rich Eu(As$_{1-x}$P$_x$)$_3$(18–21). The system undergoes two incommensurate phases upon decreasing temperature: a sinusoidal phase and, at lower temperatures, a helimagnetic phase. Both phases have almost the same magnetic propagation vector (*q* ~ (-0.726, 0, 0.255) and (-0.726, 0, 0.222), respectively(18–20)), nearly parallel to the crystalline *a*-axis. Under an applied magnetic field along the *a*-axis, approximately perpendicular to the helical plane, the magnetic structure first forms a conical structure. As the field increases, it is naturally expected that at some point the cone collapses into a fan structure(22), where the moment along the *c*-axis vanishes and leaves an oscillating component along the magnetically easy *b*-axis. Eventually, this evolution ends in a field-forced ferromagnetic (FM) configuration.

The above process is schematically shown in Fig. 1C. As can be seen, the symmetry constraints in this system forbids any energy degeneracy between an electronic state with momentum $k_q$ and spin $s_q$, $E(k_q, s_q)$, with $E(-k_q, s_q)$ or $E(k_q, -s_q)$, where $k_q$ and $s_q$ are components along the helimagnetic propagation vector $q$. This results in a symmetric spin-polarized band structure in the helimagnetic phase. Applying a magnetic field along $q$, where the conical magnetic structure is realized, causes an asymmetry in the band structure, which we later show induces nonreciprocity in the electronic transport. When the magnetic field is strong enough, the band structure becomes symmetric again with a fan-shaped achiral magnetic structure. Eventually, all spins are fully aligned with the magnetic field, leading to an FM structure. The variation of the band structure in the course of magnetic phase transitions was partly discussed in a simplified theoretical model(23) but has never been observed in a real material.

The resistivity in helimagnets along the propagation vector can be described up to the nonlinear regime as

$$\rho(j) = \rho_1 + \rho_2 j + \cdots \quad (1),$$

where $\rho_1$ ($= \rho^{1\omega}$) and $\rho_2$ reflect the ordinary resistivity and nonlinear resistivity, respectively, and $j$ is the electric current density. The second term can be sensitively measured as a second-harmonic voltage under an ac electric current $j = j_0 \sin\omega t$(8);

$$E = \rho^{1\omega} j_0 \sin\omega t + \rho^{2\omega} j_0 (1 - \cos 2\omega t) + \cdots \quad (2).$$

$\rho^{2\omega} = \rho_2 j_0/2$ is composed of field-symmetric and field-asymmetric components as $\rho^{2\omega} = \rho^{2\omega}_{sym} + \rho^{2\omega}_{asym}$, where $\rho^{2\omega}_{sym}(+H) = \rho^{2\omega}_{sym}(-H)$ and $\rho^{2\omega}_{asym}(+H) = -\rho^{2\omega}_{asym}(-H)$. $\rho^{2\omega}_{asym}$ is caused by the nonreciprocal electronic transport while $\rho^{2\omega}_{sym}$ may be induced by some extrinsic effect such as the effect of electrode contact. Because $\rho^{2\omega} \propto j_0$, a large electric current density $j_0$ is preferred for the measurement of nonlinear electronic transport. To realize a large current density, the α-EuP$_3$ single crystal sample was microfabricated into a device with a cross-sectional area of ~ 8 × 4 μm$^2$ (Fig. 2A), using the focused ion-beam (FIB) technique(24). The temperature dependence of resistivity for the FIB sample clearly shows two magnetic transitions at low temperatures (Fig. 2B). The overall resistivity behavior is consistent with previously reported bulk sample results(14), assuring the device quality.

Fig. 2C exemplifies the second-harmonic resistivity $\rho^{2\omega}$ as a function of the magnetic field at 4.2 K, measured by a lock-in amplifier. $\rho^{2\omega}$ shows an evident field-asymmetric behavior with minimal hysteresis. $\rho^{2\omega}$ seems mostly composed of an intrinsic nonreciprocal electronic resistivity $\rho^{2\omega}_{asym}$, which is induced by the finite chirality of the system arising from helimagnetic order. Fig. 2D shows the field asymmetric component $\rho^{2\omega}_{asym} = (\rho^{2\omega}_{0 \to +1T}(+H) - \rho^{2\omega}_{0 \to -1T}(-H))/2$, where the subscripts of $\rho^{2\omega}_{0 \to +1T}$ and $\rho^{2\omega}_{0 \to -1T}$ describe the sweeping direction of the magnetic field. Note that the sign and magnitude are not reproduced in each measurement. This is because there are two energetically degenerate chiral states in the centrosymmetric crystal α-EuP$_3$, and the chiral domain population differs in each measurement. Recent studies on centrosymmetric helimagnetic metals with Mn magnetic moments have reported that the helimagnetic chirality can be controlled by traversing the achiral-chiral transition while simultaneously applying a magnetic field and a high-density dc electric current(6, 7, 25, 26). The sign of the controlled chirality depends on whether the magnetic field and dc current are parallel or antiparallel. Following these reports, we have attempted to control the chirality by a similar procedure with the magnetic field $H_p$ and dc current $J_p$ before the measurement of $\rho^{2\omega}$ shown in Figs. 2C and 2D (for the details of the procedure, see the Supporting Information). As shown in Fig. 2D, the sign of $\rho^{2\omega}_{asym}$ does not depend on $H_p$ and $J_p$ but seems random, indicating the control procedure was not effective in α-EuP$_3$. Nevertheless, the magnetic field variations were generally reproducible. Notably, $\rho^{2\omega}_{asym}$ shows a complex magnetic field dependence. As the magnetic field is increased from 0 T, it first shows a broad peak and then steeply changes accompanying a sign change followed by a sharp peak. Fig. 2E shows the magnetic field dependence of $\rho^{2\omega}_{asym}$ measured at temperatures of 4.2 K, 6.1 K, 8.0 K, and 10.0 K. The $\rho^{2\omega}$ measurements were done several times at each temperature, and the presented data correspond to the maximum response obtained. $\rho^{2\omega}_{asym}$ distinctly demonstrates nonreciprocity only within the chiral magnetic phase (4.2 K and 6.1 K) and vanishes in the achiral phases (8.0 K and 10.0 K). These results clearly show the correspondence between $\rho^{2\omega}_{asym}$ and the magnetic symmetry.

Let us now examine the detailed magnetic field dependence of the $\rho^{2\omega}_{\text{asym}}$ signal across chiral-achiral transitions. Fig. 3D displays the nonreciprocal electronic response at 4.2 K while sweeping the magnetic field from +4 T to -4 T without any chitrality-control procedure, with the data antisymmetrized as $\rho^{2\omega}_{\text{asym}} = [\rho^{2\omega}(+H) - \rho^{2\omega}(-H)]/2$. This antisymmetrization is valid when the magnetic hysteresis is negligible, as confirmed in Fig. 2C. The obtained data is scaled to that shown in Fig. 2E by multiplying a constant. The $\rho^{2\omega}_{\text{asym}}$ signal shows a sharp peak around $B_1 \sim 0.7$ T. In the high field region, $\rho^{2\omega}_{\text{asym}}$ gradually decreases and vanishes at around $B_2 \sim 2.5$ T. For comparison, we plot the magnetic field dependences of magnetization $M$ and its field derivative, longitudinal resistivity $\rho_{xx}$, and Hall resistivity $\rho_{yx}$ in Figs. 3A-C. Clear anomalies at $B_2$ and $B_3 \sim 3.1$ T correspond to the transition from conical to fan and from fan to field-forced FM phases, respectively. The disappearance of $\rho^{2\omega}_{\text{asym}}$ at $B_2$ is again in excellent agreement with the magnetic symmetry (Fig. 1B). In addition, a slight anomaly is discerned in the derivative of magnetization around $B_1$. In contrast, this $B_1$ anomaly is conspicuously observed in the longitudinal and Hall resistivities, both showing a peak. In a previous paper, Mayo *et al.*(14) reported a Fermi surface reconstruction induced by the FM polarization of Eu 4*f* moments, occurring at an approximate threshold magnetization of $M \sim 1.8$ $\mu_B$/Eu. The observed anomalies at $B_1$ indeed correspond to a similar magnetization as the reported threshold. The fact that the anomaly is notable in the transport properties and less visible in magnetization suggests a strong role of the Fermi surface in the observed nonreciprocity (for further notes on magnetotransport results, see the Supporting Information).

To understand how the experimentally observed nonreciprocity in α-EuP$_3$ arises from microscopic changes in its band structure, we have conducted first-principles band calculations for the relevant magnetic structures; helical, conical, fan, and FM (see Methods and Supporting Information). We construct each magnetic phase by appropriately constraining the directions of Eu magnetic moments in a sufficiently large supercell as shown in Supplementary Fig. S8. It is to be noted that the reported helical magnetic order is incommensurate and slightly tilted from the crystalline *a*-axis ($q$ ~ (-0.726, 0, 0.255)). We have instead considered a commensurate helimagnetic model with $q$ = (-0.75, 0, 0) and a helical plane perpendicular to $q$ for the sake of simplicity. To model the intermediate phases induced by the magnetic field, the Eu magnetic moments were gradually tilted along the *a*-axis through a systematic linear interpolation between the helimagnetic and ferromagnetic phase. This resulted in a net magnetization along the *a*-axis with a conical magnetic configuration, which we used to find its corresponding magnetic field in our experimental setup $\mu_0 H$ // $a$ in this phase. The fan structure was separately constructed by setting the $c^*$-axis ($c^*$ // $a \times b$) components of the magnetic moments to zero and collapsing the conical structure. Fig. 4A displays the band structure calculated for each of these magnetic phases along the -Y – Γ – +Y direction, corresponding to the $a^*$-axis direction ($a^*$ // $b \times c$). As can be seen, in the helical phase ($M$ = 0 $\mu_B$/Eu), the spin degeneracy is lifted due to a spin dependent shift along the Y direction rather than a Zeeman-like splitting, characteristic to a $\mathcal{P}$-broken system. It should be noted that the momentum dependent spin splitting seen here differs from the Rashba-type, in which the shifted direction in momentum space is perpendicular to the spin direction. Rather, the spin direction of the split bands is parallel to the direction of momentum shift, similar to the case of chiral materials such as Te(27). Nonetheless, the overall band structure is symmetric along -Y – Γ – +Y.

Upon introducing a conical deformation in the magnetic structure, achieved through applying a relatively weak magnetization along the *a*-axis, a notable transformation unfolds within the band structure, evident as a pronounced asymmetry in the band dispersions along -Y – Γ – +Y. This asymmetry implies a discernible distinction in the charge current induced by an electric field +*E* as compared to that of -*E*. Consequently, the system is anticipated to manifest a finite nonreciprocal electronic resistivity within the conical phase. Remarkably, as the magnetic phase transitions away from the conical phase into the achiral fan magnetic or ferromagnetic phase, a restoration of symmetry becomes apparent in the band structure along -Y – Γ – +Y, as illustrated in Fig. 1C. This intriguing finding provides compelling evidence as to why nonreciprocal electronic transport is exclusively observable in the conical magnetic structure and explains how this distinct asymmetry in the band dispersions is responsible for it.

To further discuss the effect of field variation in the conical magnetic state, we turn our attention to the conical angle dependence of the band structure. Fig. 4B schematically illustrates the conical angle dependent evolution of the relevant Fermi pockets. Here we show the pockets corresponding to the energy level indicated by the dashed lines in Fig. 4A. As can be seen they appear as two symmetric pockets in the helimagnetic state, as expected. However, upon tilting the magnetic moments along the $a$-axis and, hence, increasing the conical angle, an asymmetric deformation of the energy pockets emerges. In the FM state, the energy pockets regain symmetry.

Another notable feature seen in the experiment is a sharp sign change followed by a negative peak in $\rho^{2\omega}_{asym}$ under increasing magnetization (Fig. 4C). As a measure of band asymmetry, we plot the residual density of states, defined as the difference between the total density of states for $+k_{a*}$ and that obtained for $-k_{a*}$, in the conical phase as a function of conical magnetization in Fig. 4D. As can be seen, it initially shows a positive value followed by a sign change and a discernible negative peak. This behavior is consistent with the experimental observation of the nonreciprocal resistivity (Fig. 4C). Despite its apparent simplicity, it is remarkable that our theoretical model not only captures but also qualitatively agrees with the experiment, further providing compelling evidence that the nonreciprocal electronic transport is induced by the asymmetry in the momentum-space electronic bands.

In summary, we have investigated nonreciprocal electronic transport in a magnetic semimetal α-EuP$_3$. The material facilitates successive tunability of the magnetic structure from helimagnetic to induced ferromagnetic structures via conical and fan structures in response to a magnetic field, presenting a rich array of transport responses coupled with the magnetic phases. The nonreciprocal electronic transport was found to arise as the chiral magnetic structure imposes an asymmetric deformation in the electronic band structure, and vanish as the magnetic structure is tuned into an achiral phase, restoring symmetry in momentum space. Moreover, a singular sign change of nonreciprocal electronic transport in the middle of the conical phase is observed, seemingly caused by the reversal of band asymmetry in the course of the dramatic exchange splitting inherent in this material. These results unravel that the nonreciprocal electronic transport in helimagnets originates from the band asymmetry caused by the magnetic chirality. Our findings suggest that the understanding of the electronic band structure in the momentum space is indispensable also for designing the novel functionality in the chiral magnetic textures.

## Materials and Methods

### Single crystal growth and crystal identification

Single crystals of α-EuP$_3$ were grown by the high-pressure synthesis method(14). The crystal was identified as α-EuP$_3$ by powder and single crystal X-ray diffraction techniques.

### FIB-microfabrication of transport devices

The Hall bar device was fabricated using the focused ion-beam (FIB) technique(24) in the following steps: First, a small piece of single crystal α-EuP$_3$ with an approximate size of $100 \times 50 \times 8$ μm³ was placed on an oxidized single-crystal silicon wafer. Next, 20 nm of Pt was deposited by RF magnetron sputtering, followed by 200 nm of Au deposited by electron beam evaporation, both covering the entire sample. The sample was then fixed to the substrate by FIB-assisted Pt deposition, ensuring electrical contact between the sample and substrate electrodes, and then was patterned into a Hall bar by FIB. Finally, the Pt/Au layer on the central top surface of the Hall bar channel was etched away by FIB.

### Magnetotransport measurements

For the bulk measurements, electrode contacts were made by depositing 200 nm of Au on the sample and connecting Au wires using silver paste (see Supplementary Figure S4). Both bulk and FIB samples were subjected to transport measurements in a superconducting magnet. The ac resistivities were measured using the four-wire configuration and the standard lock-in technique, applying an ac electric current with a frequency of 11.15 Hz. The ac electric current density applied for the second-harmonic resistivity measurement in the FIB sample was set to an amplitude of $1 \times 10^7$ A/m$^2$.

### Band structure calculations

The electronic structure calculations were performed within density functional theory (DFT) using the Perdew-Burke-Ernzerhof (PBE) exchange-correlation functional(28) as implemented in the VASP program(29). To properly treat the strong on-site Coulomb interaction of Eu-4*f* states, an effective Hubbard-like potential term $U_{\text{eff}}$ was added to the PBE functional. The value of $U_{\text{eff}}$ was fixed at 6 eV for Eu-4$f$ orbitals(30) and zero for all other orbitals. Relativistic effects including spin-orbit coupling were fully taken into account. The cut-off energy for plane waves forming the basis set was set to 400 eV. The lattice parameters and atomic positions were taken from experiment(14). The corresponding Brillouin zone for the supercell calculation was sampled using a $2 \times 8 \times 8$ $k$-mesh. The calculation for each magnetic structure was carried out by imposing the corresponding magnetic moment configurations on the Eu site. The density-of-states calculations were performed using a 64-band effective Hamiltonian, downfolded from the DFT calculations using maximally localized Wannier functions(31). The sampling of the corresponding Brillouin zone was done using a $240 \times 240 \times 240$ $k$-mesh.

**Acknowledgments**

The authors thank Y. Kodama and T. J. Konno for providing thorough technical assistance in using the FIB technique, and T. Taniguchi and M. Fujita for the single crystal x-ray measurements. The authors also acknowledge Y. Fujishiro, H. Sakai, and M. Kimata for their insightful advice regarding the device fabrication process. This study was supported in part by Japan Society for the Promotion of Science KAKENHI (Grant No. 21H01036, 22H04461, 22J01078, 22KJ0212, 22H00343, 23H04871, 23K13654). This study was partially supported by the Center for Integrated Nanotechnology Support (CINTS) at Tohoku University and also by the Advanced Research Infrastructure for Materials and Nanotechnology (ARIM) project of the Ministry of Education, Culture, Sports, Science and Technology (MEXT), Japan. The authors gratefully acknowledge the Center for Computational Materials Science at the Institute for Materials Research for allocations on the MASAMUNE- IMR supercomputer system (Project no. 202112-SCKXX-0510). M.S.B. acknowledges support from Leverhulme Trust (Grant No. RPG-2023-253). D.-A.D. had support from EPSRC Standard Research Studentship (DTP) EP/T517823/1.

# References

1. L. D. Barron, From cosmic chirality to protein structure: Lord Kelvin's legacy. *Chirality* **24**, 879–893 (2012).

2. Y. Tokura, N. Kanazawa, Magnetic Skyrmion Materials. *Chem. Rev.* **121**, 2857–2897 (2021).

3. K. Shimizu, S. Okumura, Y. Kato, Y. Motome, Spin moiré engineering of topological magnetism and emergent electromagnetic fields. *Phys. Rev. B* **103**, 184421 (2021).

4. T. Yokouchi, *et al.*, Electrical magnetochiral effect induced by chiral spin fluctuations. *Nat. Commun.* **8**, 866 (2017).

5. R. Aoki, Y. Kousaka, Y. Togawa, Anomalous Nonreciprocal Electrical Transport on Chiral Magnetic Order. *Phys. Rev. Lett.* **122**, 057206 (2019).

6. N. Jiang, Y. Nii, H. Arisawa, E. Saitoh, Y. Onose, Electric current control of spin helicity in an itinerant helimagnet. *Nat. Commun.* **11**, 1601 (2020).

7. H. Masuda, *et al.*, Room temperature chirality switching and detection in a helimagnetic $MnAu_2$ thin film. *Nat. Commun.* **15**, 1999 (2024).

8. Y. Tokura, N. Nagaosa, Nonreciprocal responses from non-centrosymmetric quantum materials. *Nat. Commun.* **9**, 3740 (2018).

9. G. L. Rikken, J. Fölling, P. Wyder, Electrical magnetochiral anisotropy. *Phys. Rev. Lett.* **87**, 236602 (2001).

10. G. L. J. A. Rikken, P. Wyder, Magnetoelectric anisotropy in diffusive transport. *Phys. Rev. Lett.* **94**, 016601 (2005).

11. V. Krstić, S. Roth, M. Burghard, K. Kern, G. L. J. A. Rikken, Magneto-chiral anisotropy in charge transport through single-walled carbon nanotubes. *J. Chem. Phys.* **117**, 11315–11319 (2002).

12. F. Pop, P. Auban-Senzier, E. Canadell, G. L. J. A. Rikken, N. Avarvari, Electrical magnetochiral anisotropy in a bulk chiral molecular conductor. *Nat. Commun.* **5**, 3757 (2014).

13. T. Ideue, *et al.*, Bulk rectification effect in a polar semiconductor. *Nat. Phys.* **13**, 578–583 (2017).

14. A. H. Mayo, *et al.*, Magnetic Generation and Switching of Topological Quantum Phases in a Trivial Semimetal α−$EuP_3$. *Phys. Rev. X.* **12** (2022).

15. M. A. Ruderman, C. Kittel, Indirect Exchange Coupling of Nuclear Magnetic Moments by Conduction Electrons. *Phys. Rev.* **96**, 99–102 (1954).

16. T. Kasuya, A Theory of Metallic Ferro- and Antiferromagnetism on Zener's Model. *Prog. Theor. Phys.* **16**, 45–57 (1956).

17. K. Yosida, Magnetic Properties of Cu-Mn Alloys. *Phys. Rev.* **106**, 893–898 (1957).

18. P. J. Brown, T. Chattopadhyay, The helimagnetic structure of Eu(As$_{0.20}$P$_{0.80}$)$_3$ determined by zero-field neutron polarimetry. *J. Phys. Condens. Matter* **9**, 9167 (1997).

19. T. Chattopadhyay, P. J. Brown, H. G. von Schnering, Sine-wave---to---helimagnetic transition in phosphorus-rich Eu(As$_{1-x}$P$_x$)$_3$. *Phys. Rev. B* **36**, 7300–7302 (1987).

20. T. Chattopadhyay, P. J. Brown, P. Thalmeier, W. Bauhofer, H. G. von Schnering, Neutron-diffraction study of the magnetic ordering in EuAs$_3$, Eu(As$_{1-x}$P$_x$)$_3$, and β-EuP$_3$. *Phys. Rev. B* **37**, 269–282 (1988).

21. Y. Ōnuki, *et al.*, Magnetic and Fermi Surface Properties of Semimetals EuAs$_3$ and Eu(As$_{1-x}$P$_x$)$_3$. *J. Phys. Soc. Jpn.* **92**, 114703 (2023).

22. J. Jensen, A. R. Mackintosh, *Rare Earth Magnetism* (Clarendon Press, Oxford, 1991).

23. A. Zadorozhnyi, Y. Dahnovsky, Anomalous Hall effect in conical helimagnetic crystals. *Phys. Rev. B* **107**, 035202 (2023).

24. P. J. W. Moll, Focused Ion Beam Microstructuring of Quantum Matter. *Annu. Rev. Condens. Matter Phys.* **9**, 147–162 (2018).

25. N. Jiang, *et al.*, Chirality Memory Stored in Magnetic Domain Walls in the Ferromagnetic State of MnP. *Phys. Rev. Lett.* **126**, 177205 (2021).

26. J.-I. Ohe, Y. Onose, Chirality control of the spin structure in monoaxial helimagnets by charge current. *Appl. Phys. Lett.* **118**, 042407 (2021).

27. M. Sakano, *et al.*, Radial Spin Texture in Elemental Tellurium with Chiral Crystal Structure. *Phys. Rev. Lett.* **124**, 136404 (2020).

28. J. P. Perdew, K. Burke, M. Ernzerhof, Generalized Gradient Approximation Made Simple. *Phys. Rev. Lett.* **77**, 3865–3868 (1996).

29. Vienna Ab Initio Simulation Package, Version 5.3.5 (2014).

30. A. S. Borukhovich, A. V. Troshin, *Europium Monoxide: Semiconductor and Ferromagnet for Spintronics* (Springer, 2018).

31. A. A. Mostofi, *et al.*, An updated version of wannier90: A tool for obtaining maximally-localised Wannier functions. *Comput. Phys. Commun.* **185**, 2309–2310 (2014).

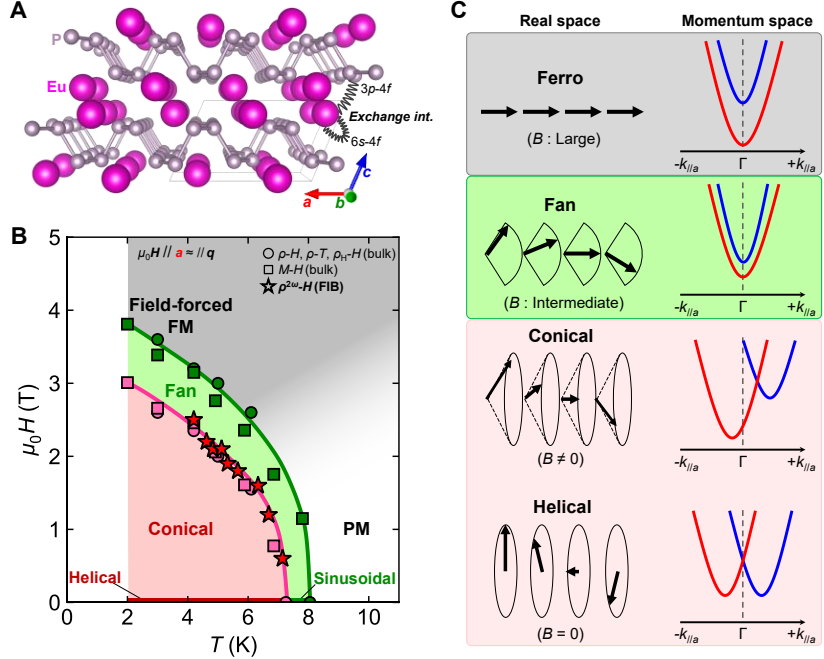

**Figure 1. Crystal, magnetic, and expected band structures in α-EuP$_3$.**

(**A**) Crystal structure of α-EuP$_3$. The localized Eu-4$f$ moments couple with the itinerant electrons in the system, which consist of P-3$p$ and Eu-6$s$ orbitals. (**B**) Field-temperature phase diagram of α-EuP$_3$ for a magnetic field $\mu_0 H \,/\!/\, a$, approximately aligned with the propagation vector $q$ of the magnetic phases. (**C**) Schematic illustrations of the symmetrically expected relationship between the real-space magnetic structure and the momentum-space band structure. Although the band structures are symmetric under helical, fan, and ferromagnetic structures, the conical magnetic structure induced by a moderate magnetic field applied along the $q$-vector gives rise to an asymmetrical deformation of the band structure along the $k$-path in the $q$-vector direction.

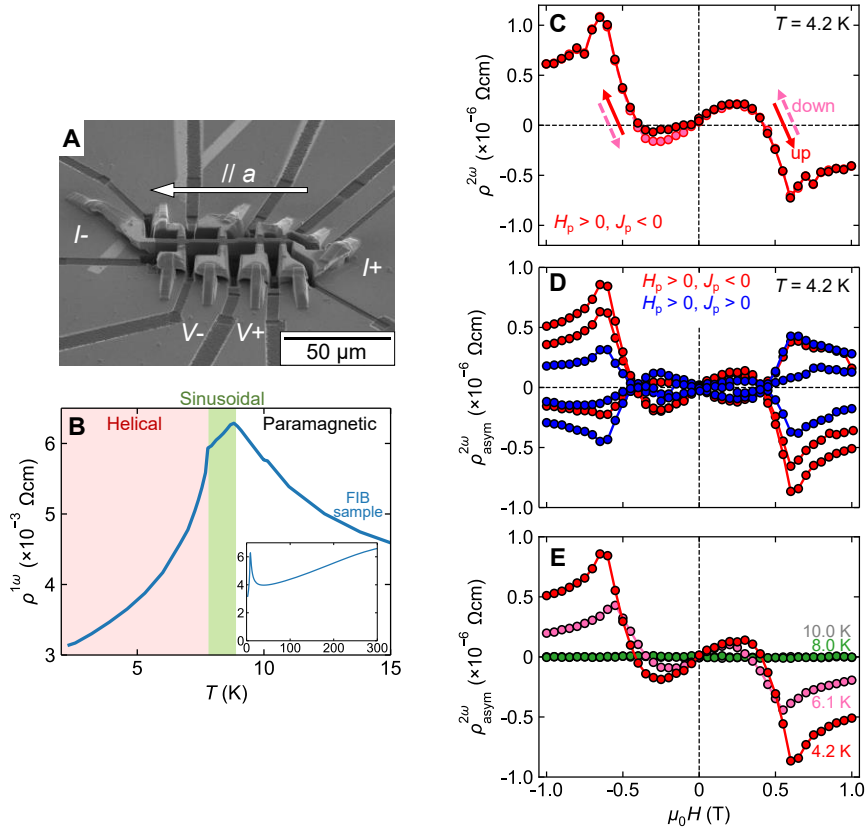

**Figure 2. Nonreciprocal electronic transport measured in a microfabricated α-EuP₃ sample.**

(**A**) Scanning electron microscope (SEM) image of the microfabricated α-EuP₃ sample by the FIB technique. (**B**) Temperature dependence of resistivity near the magnetic transition temperatures for the microfabricated sample. The inset shows the resistivity data up to room temperature. (**C**) Magnetic field dependence of the second-harmonic resistivity $\rho^{2\omega}$ after the attempt of chirality control with anti-parallel magnetic field and electric current ($H_p > 0$, $J_p < 0$). The red (pink) marker indicates the field increase (decrease) sweep. (**D**) Field-asymmetric component of the second-harmonic resistivity $\rho^{2\omega}_{asym}$ measured after attempting the chirality control with the magnetic field $H_p$ and electric current $J_p$. The blue and red markers indicate the parallel and anti-parallel $H_p$ and $J_p$, respectively. (**E**) Temperature dependence of the maximum field-asymmetric component of the second-harmonic resistivity $\rho^{2\omega}_{asym}$.

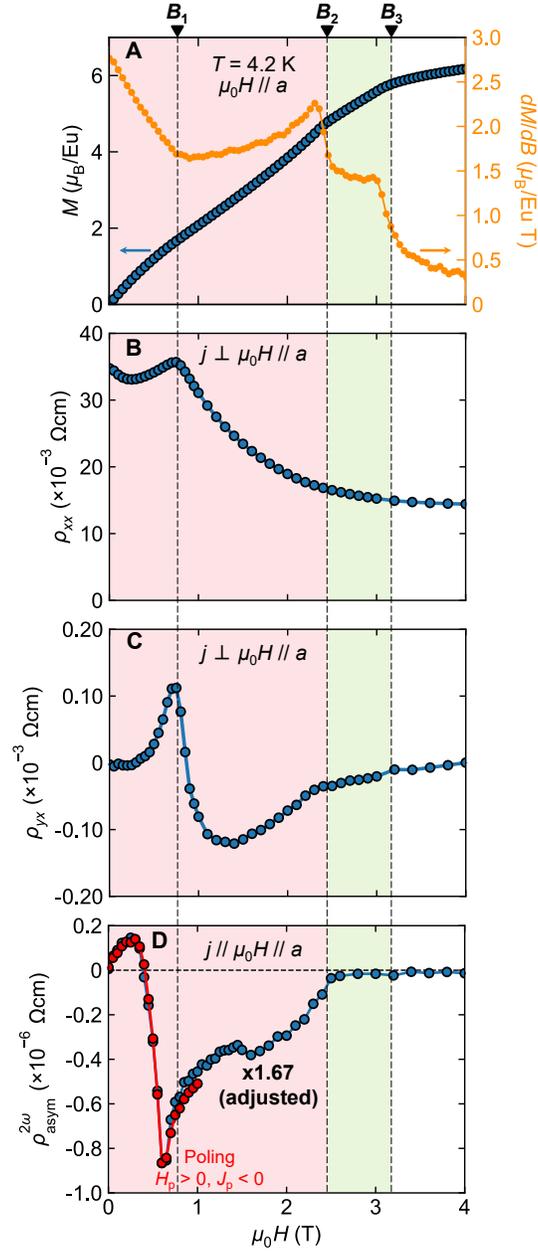

**Figure 3. Magnetic field dependence of transport and magnetic properties across chiral-achiral phase transitions.**

Magnetic field dependence of the (**A**) magnetization $M$ and its field derivative, (**B**) longitudinal resistivity $\rho_{xx}$, (**C**) transverse (Hall) resistivity $\rho_{yx}$, and (**D**) $\rho^{2\omega}_{\text{asym}}$, at $T = 4.2$ K under $\mu_0 H \mathbin{/\mkern-6mu/} a$ up to 4 T. (**A**-**C**) were measured in bulk samples and (**D**) was measured in the FIB sample. $\rho^{2\omega}_{\text{asym}}$ shown in blue in panel **D** was measured without attempting chirality control. The magnitude is adjusted to match the maximum value presented in Fig. 2D.

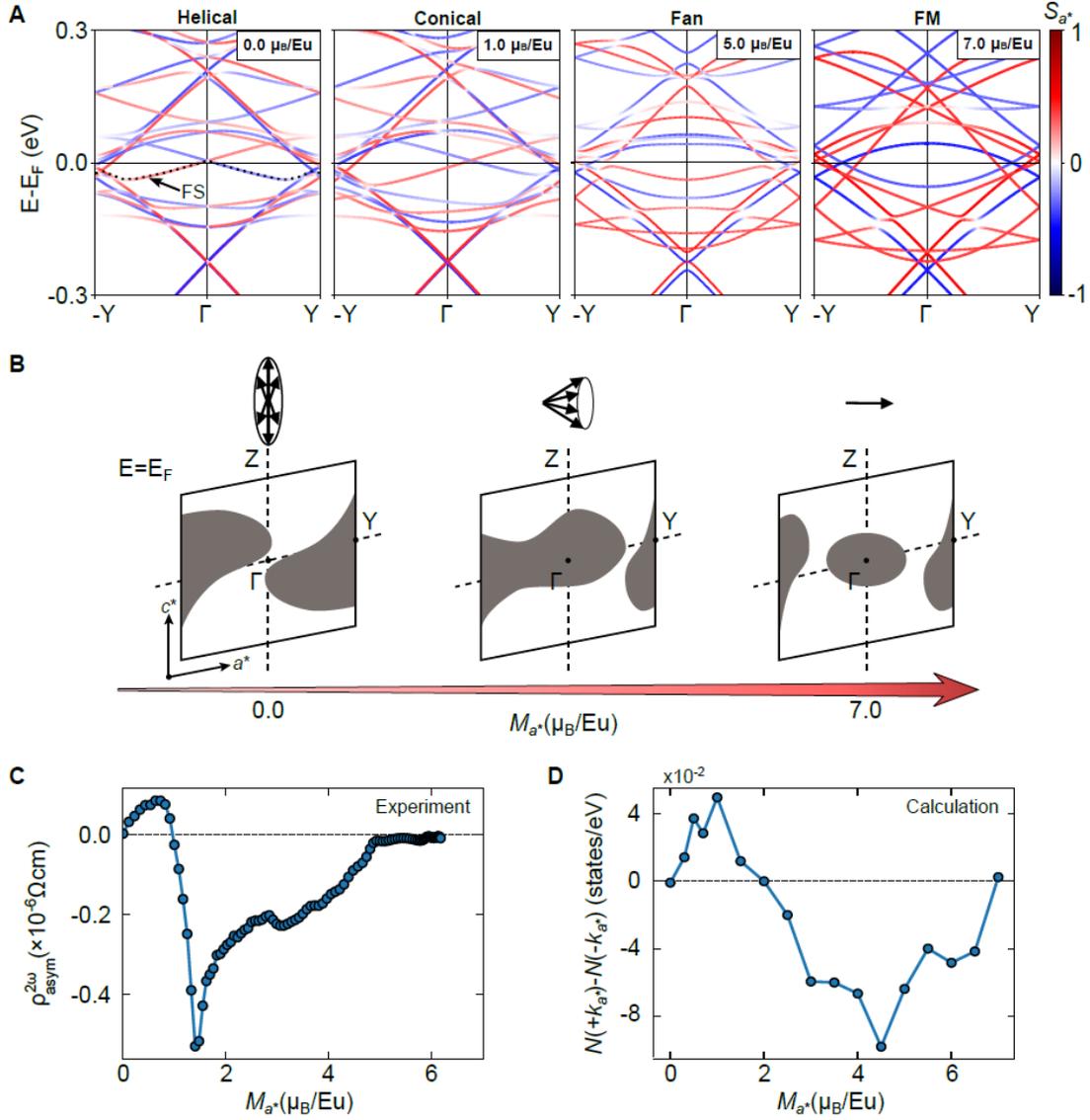

**Figure 4. Electronic band structures of α-EuP$_3$ in each magnetically ordered state.**

(**A**) The electronic band structures in the helimagnetic, conical, fan, and field-forced ferromagnetic (FM) states, respectively. The calculation was carried out by tilting the Eu magnetic moments along the *a*-axis and imposing the magnetic structures corresponding to the experimental setup of $\mu_0 \boldsymbol{H}$ // $\boldsymbol{a}$. (**B**) Schematic of the field evolution of the Fermi pockets derived from a single electron band denoted in panel **A** in the conical magnetic state. (**C**) Magnetization dependence of experimental $\rho^{2\omega}_{\text{asym}}$. (**D**) Magnetization dependence of the difference in the calculated density of states between $+k_{a^*}$ and $-k_{a^*}$ in the conical magnetic state.

# Supporting Information for
# Band-asymmetry-driven nonreciprocal electronic transport in a helimagnetic semimetal α-EuP$_3$

Alex Hiro Mayo[1]*, Darius-Alexandru Deaconu[2], Hidetoshi Masuda[1], Yoichi Nii[1], Hidefumi Takahashi[3], Rodion Vladimirovich Belosludov[1], Shintaro Ishiwata[3], Mohammad Saeed Bahramy[2], and Yoshinori Onose[1]

[1]Institute for Materials Research, Tohoku University; Aoba-ku, Sendai, 980-8577, Japan.

[2]Department of Physics and Astronomy, School of Natural Sciences, The University of Manchester, Oxford Road, Manchester M13 9PL, United Kingdom

[3]Division of Materials Physics, Graduate School of Engineering Science, Osaka University, Toyonaka, Osaka 560-8531, Japan.

*Corresponding author. Email: alex.hiro.mayo.d1@tohoku.ac.jp

## Section 1. Multiple trials of the chirality-control procedure

Figure S1 schematically illustrates the relation between the chirality-control procedure and the magnetic phases in α-EuP$_3$.

In a centrosymmetric helimagnet, the opposing chirality are energetically degenerate and therefore the system will basically form magnetic domains of opposite chirality when undergoing helimagnetic ordering. In a recent study, Jiang *et al.*[6] reported that the helimagnetic chirality in a centrosymmetric crystal can be controlled by traversing the achiral-chiral transition while simultaneously applying a magnetic field and a high-density dc electric current, which they referred to as the "poling" procedure. The simultaneous application of the magnetic field and the electric current lifts the degeneracy relevant to the chirality and energetically favors one chirality over the other, depending on whether the magnetic field and dc current are parallel or antiparallel. A sufficiently large electric current density is essential for this procedure; thus, microfabrication of the sample is again vitally important. Jiang *et al.* achieved chirality-control in centrosymmetric MnP, controlling the system into a single chirality domain. A constant dc electric current density in the order of $1\times10^9$ A/m$^2$ was applied while sweeping the magnetic field, traversing the achiral-chiral phase boundary. It was also confirmed that the chirality can be switched to the opposite sign depending on the parallel/antiparallel application of the magnetic field and electrical current. The effectiveness of the poling procedure was also demonstrated in another Mn-based centrosymmetric metal MnAu$_2$[7].

Following these reports, we have attempted to control the chirality by a similar procedure with the magnetic field $H_\mathrm{p}$ and dc current $J_\mathrm{p}$ before the measurement of $\rho^{2\omega}$. Because the current induced heating is significant at the low temperature region, we utilized the sample's Joule heating to traverse the magnetic phases, as an alternative to sweeping the chamber temperature. Figure S1 shows how the system traverses through magnetic phases during the poling procedure. A magnetic field of $H_p = 1.5$ T was first applied. Then, the magnitude of the dc electric current density was slowly applied up to $J_p = \pm 1\times 10^8$ A/m$^2$, traversing the magnetic phases and resulting in the paramagnetic (PM) phase due to Joule heating. Then, after slowly sweeping the dc current back down to zero, undergoing the magnetic transitions from PM to fan and fan to conical, successively, the magnetic field was quenched. This procedure with $H_p > 0$, $J_p < 0$ was executed prior to the measurement presented in Fig. 2C. Same measurements were repeated three times each with $H_p > 0$, $J_p < 0$ (antiparallel poling, shown in red) and with $H_p > 0$, $J_p > 0$ (parallel poling, shown in blue), presented in Figure 2D. The grey arrows in Figure S1 describe the path the system undergoes during the poling procedure.

Figure S2 shows the multiple trials of the poling procedure followed by a second-harmonic resistivity measurement. Same measurements were repeated three times each with $H_p > 0$, $J_p < 0$ (antiparallel poling, shown in red) and with $H_p > 0$, $J_p > 0$ (parallel poling, shown in blue) for each temperature ($T = 4.2$ K, 6.1 K, 8.0 K, 10.0 K). In Figure 2E in the main text, the data showing a large response in each of the following panels is selected and displayed.

In stark contrast to the manganese helimagnets[6, 7], the control with magnetic fields and electric current appears to be ineffective in α-EuP$_3$. One possible reason for this is the small magnitude of $J_p$. $|J_p| = 1\times 10^8$ A/m$^2$ applied here is close to the minimum value for successful chirality-control in the case of manganese helimagnets[6, 7]. Another reason can be related to the orbital character of the Fermi surface. In the manganese metals, Mn 3$d$ orbitals have a large density of states contributing to the Fermi surface[S1], whereas in α-EuP$_3$, the P-3$p$ orbitals at the Fermi surface and the Eu-4$f$ states are only weakly hybridized[14], possibly resulting in a weaker momentum-transfer between the electric current and the local moments during the poling procedure[22, S2, S3, S4]. Although in theory one can expect that a larger $|J_p|$ should overcome the critical value and enable chirality control, the sample temperature readily exceeds the Néel temperature from Joule heating, limiting the parameter window to achieve controllability in α-EuP$_3$.

## Section 2. Temperature dependence of $\rho^{2\omega}_{asym} - H$ and reproducibility of the nonreciprocal response in a different FIB device

Figure S3 displays $\rho^{2\omega}_{asym}$ measured in a different FIB sample from the one discussed in the main text. The device is shown in panel **A**. The electrode contacts were made by spot welding Au wires on a crystal. The crystal was then patterned into a Hall bar with a cross-sectional area of ~ 40 × 6 μm² by using the FIB technique. The temperature and magnetic field dependencies are quite similar to those presented in the main text, which confirms the reproducibility. The results from both samples were used to plot the transition fields and the characteristic fields in Figure 1B and Figure S7, respectively.

**Section 3. Description of the bulk measurements**

Figure S4 shows the crystal orientation and measurement configuration for the bulk transport measurements presented in the main text. The crystal orientation is displayed in panels **A** and **B**. Electrode contacts were made by depositing 200 nm of Au on the sample. The sample was fixed on an oxidized single-crystal silicon wafer using varnish and Au wires were connected to the Au electrodes using silver paste, as shown in panel **C**. The magnetic field was applied perpendicular to the sample, parallel to the crystalline *a*-axis.

Figures S5 and S6 show isotherms of the magnetic field dependence of magnetization and Hall resistivity, respectively. The low-field anomaly, denoted as $B_1$ in Fig. 3, is more evidently observed in the Hall effect rather than magnetization, indicating a Fermi-surface-related origin.

**Section 4. Relationship between resistivity anomalies and magnetic phases**

Figure S7 shows the magnetic phase diagram under $\mu_0 H \; // \; a$ in an extended temperature region up to 20 K. In addition to the information presented in Figure 1B, the following diagram displays the low-field anomalies (shown in black) observed in Figure 3. $B_1$, $B_2$, and $B_3$ which were denoted in Fig. 3 are denoted here as well. As noted in the main text, one can see in Fig. S7 that the critical points $B_1$ and $B_2$ observed in $\rho^{2\omega}_{asym}$ agree well with the peaks or kinks also seen in the longitudinal and Hall resistivities and magnetization.

Although $\rho^{2\omega}_{asym}$ is limited only within the chiral phase, a characteristic feature about the $B_1$-anomaly in the Hall resistivity extends into the PM region. The peak structure seen in Fig. 3C, although pronounced in the magnetically ordered phase, persists even in the PM region (Fig. S6). The $B_1$-anomaly throughout the whole temperature range is shown as a grey line in Fig. S7. Notably, the anomalies shown along the grey line correspond to the magnetization of 1.7 – 1.8 $\mu_B$/Eu, which is the critical value where the Fermi surface reconstruction has been reported previously[14].

**Section 5. First-principles calculations for the magnetic phases**

Figure S8 shows the simplified models used in the electronic structure and density of states calculations presented in Figure 4.

The helimagnetic structure (Fig. S8A) was constructed based on the magnetic structure determined for isostructural P-rich Eu(As$_{1-x}$P$_x$)$_3$ ($x$ = 0.80)[18]. The structure was slightly simplified for computability by making it a commensurate structure and aligning the propagation vector parallel to the *a*-axis, with components of $\boldsymbol{q}$ = (-0.75, 0, 0).

The conical structure (Fig. S8B) was then constructed by adding a component parallel to the *a*-axis and tilting the magnetic moments, corresponding to the experimentally applied magnetic field direction. The ferromagnetic structure (Fig. S8D) is an extrapolation of the conical model and corresponds to the high-field limit where finally all the magnetic moments align parallel to the *a*-axis.

The fan structure (Fig. S8C) was constructed by setting the *c**-axis component (*c** // $\boldsymbol{a} \times \boldsymbol{b}$, i.e., along the out-of-layer direction with respect to the layered crystal structure) to zero, resulting in a net magnetic moment along the *a*-axis and an oscillating component in the *b*-axis.

## Section 6. Magnitude of the nonreciprocal electronic transport in reported polar and chiral systems

The nonlinear resistivity in $\mathscr{P}$-broken systems can be further described in detail as

$$\rho(\boldsymbol{B}, \boldsymbol{j}) = \rho_0 \left( 1 + \gamma \boldsymbol{B} \cdot \boldsymbol{j} + \cdots \right)$$

in chiral systems, and

$$\rho(\boldsymbol{B}, \boldsymbol{j}) = \rho_0 \left[ 1 + \gamma (\boldsymbol{P} \times \boldsymbol{B}) \cdot \boldsymbol{j} + \cdots \right],$$

in polar systems, where $\rho_0$ is the linear (ordinary) resistivity at zero magnetic field and the second term represents the directional nonlinear resistivity[8]. $\boldsymbol{B}$ is the magnetic flux density and $\boldsymbol{P}$ is a unit vector parallel to the polarization vector. This nonlinear nonreciprocal transport in systems with broken $\mathscr{P}$ is named magnetochiral anisotropy (MCA)[9], and its magnitude can be characterized by the value of the MCA coefficient $\gamma$. In centrosymmetric systems, $\gamma = 0$, i.e., the MCA vanishes.

Table S1 shows the comparison of the MCA magnitude among various $\mathscr{P}$-broken systems. Although the $|\gamma|$ value tends to be smaller in magnetic systems compared to materials where the crystal structure breaks $\mathscr{P}$, centrosymmetric α-EuP$_3$ exhibits MCA orders of magnitudes larger than other magnetic metals, even larger than some of the materials with $\mathscr{P}$-broken crystal structures. The exquisite coupling between the magnetic texture and the semimetallic small Fermi surface may be key in this observation, since a smaller Fermi energy will be closer in energy scale to the magnetic coupling, favorable for the large band deformation by the magnetic moments and also its detection via transport responses.

**Fig. S1.**

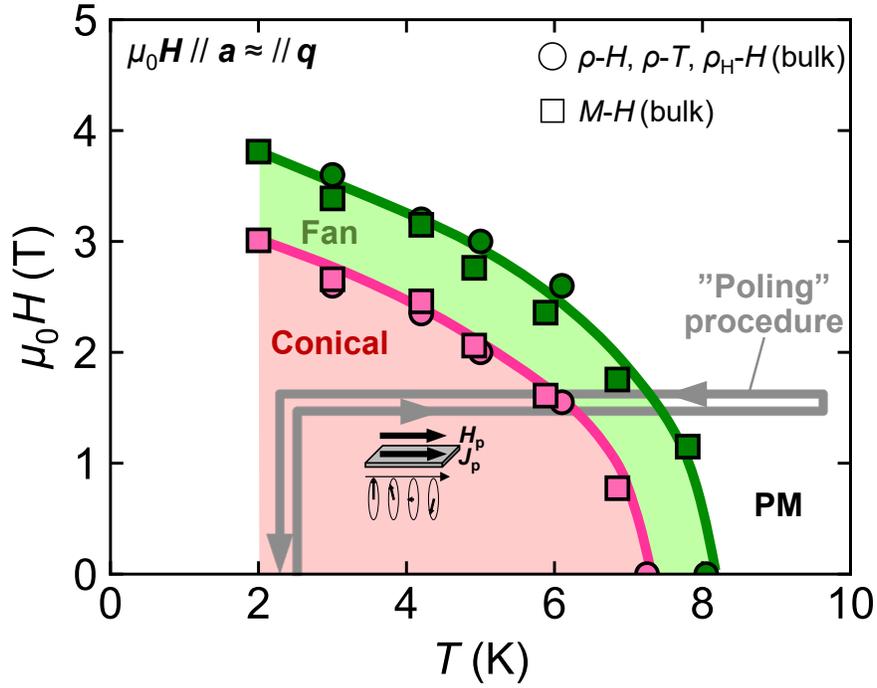

**Relation between the chirality-control procedure profile and magnetic phase diagram.** Diagram of the chirality-control (poling) procedure combining an external magnetic field $H_p$ and a dc electric current density $J_p$. The poling procedure is denoted as a grey path traversing the magnetic phases. The temperature sweep is done by sweeping the dc electric current and utilizing the Joule heating of the sample. The inset schematic illustrates the experimental configuration for the parallel poling ($H_p > 0$, $J_p > 0$) case.

**Fig. S2.**

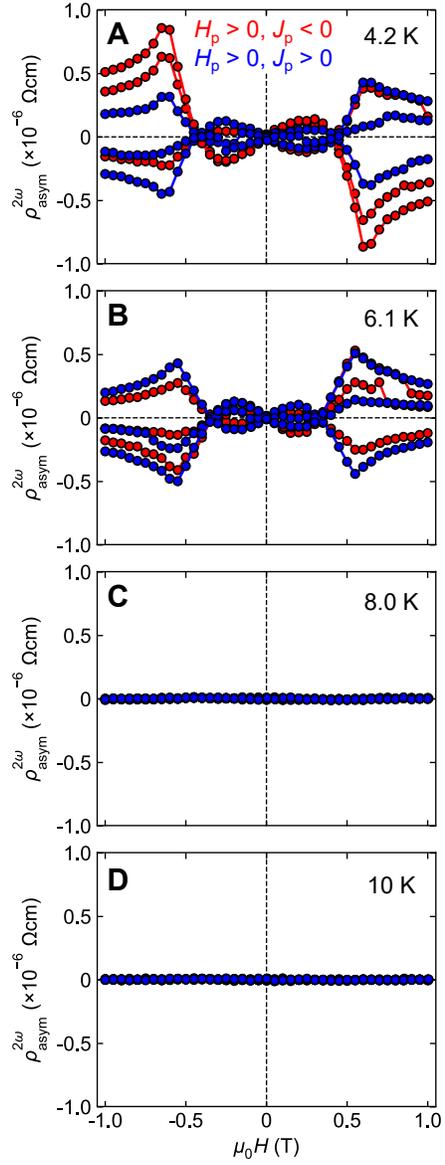

**Field-asymmetric component of the second-harmonic resistivity $\rho^{2\omega}_{asym}$ at various temperatures.** Magnetic field dependence of $\rho^{2\omega}_{asym}$ after the poling procedures at (**A**) $T = 4.2$ K, (**B**) 6.1 K, (**C**) 8.0 K, and (**D**) 10 K. The temperatures correspond to the helical/conical (4.2 K and 6.1 K), sinusoidal/fan (8.0 K), and paramagnetic (10 K) phases.

**Fig. S3.**

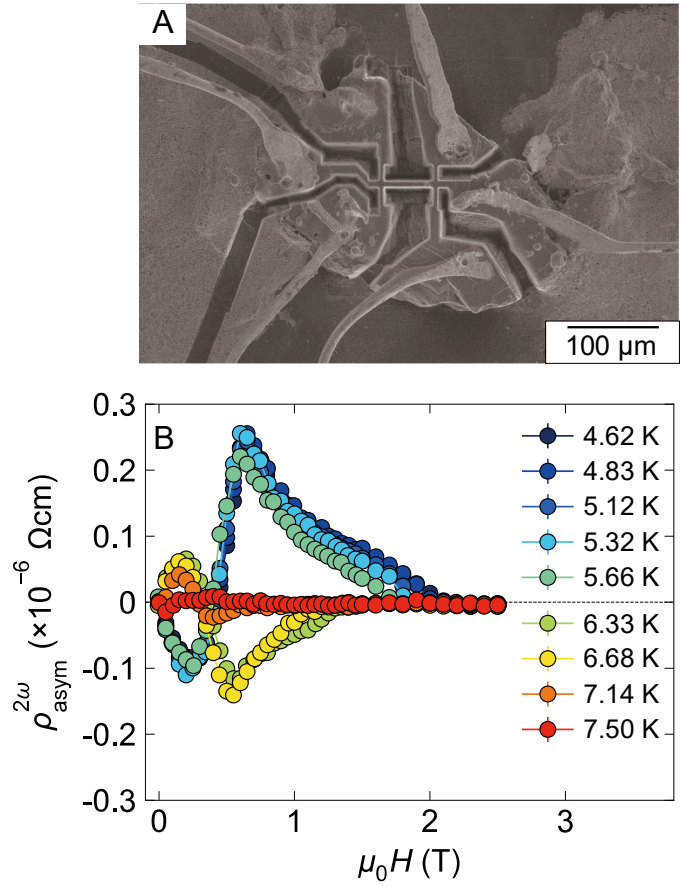

**Magnetic field dependence of $\rho^{2\omega}_{asym}$ measured in a different sample.** (**A**) SEM image of a microfabricated sample. (**B**) Isotherms of the magnetic field dependence of $\rho^{2\omega}_{asym}$ measured in the microfabricated sample shown in (**A**).

**Fig. S4.**

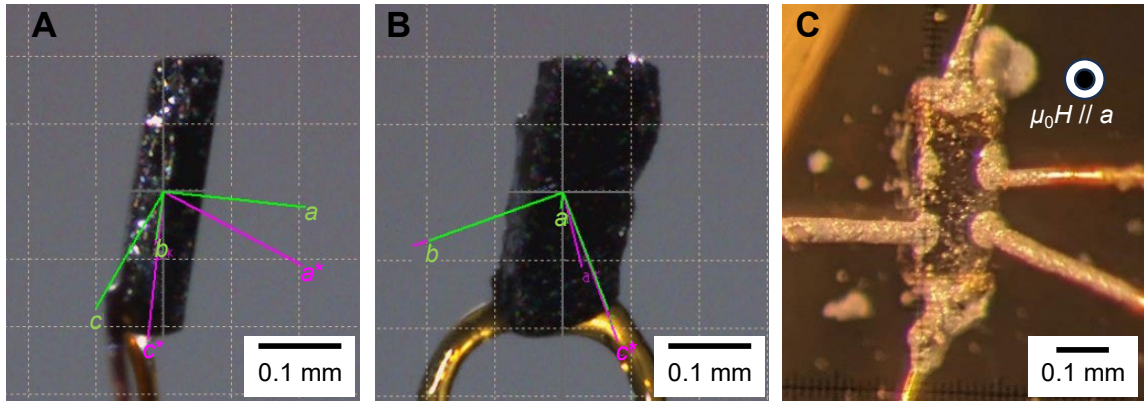

**Resistivity measurement configuration on a bulk sample.** Photographs of the sample used for bulk resistivity measurements. (**A**) and (**B**) display the crystal orientation obtained from single crystal x-ray analysis. (**C**) shows the resistivity measurement configuration.

**Fig. S5.**

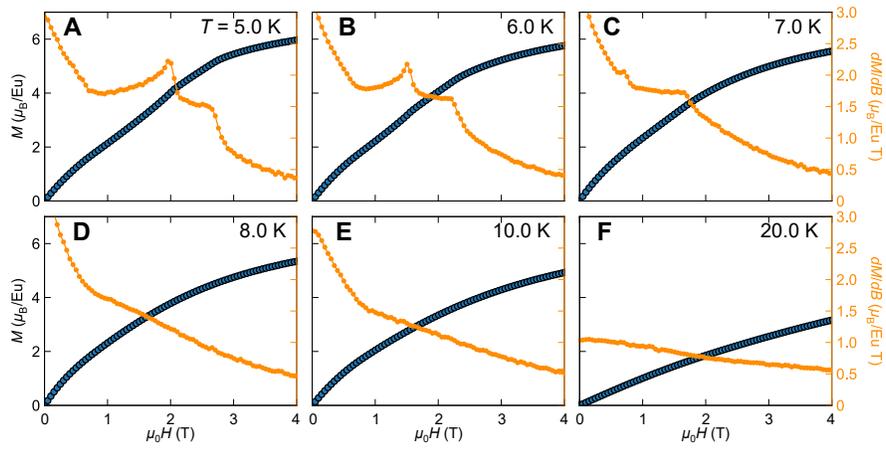

**Magnetic field dependence of magnetization.** (**A**)-(**F**) show the magnetic field dependence of the magnetization and its field derivative at various temperatures up to 20 K measured in a bulk sample.

**Fig. S6.**

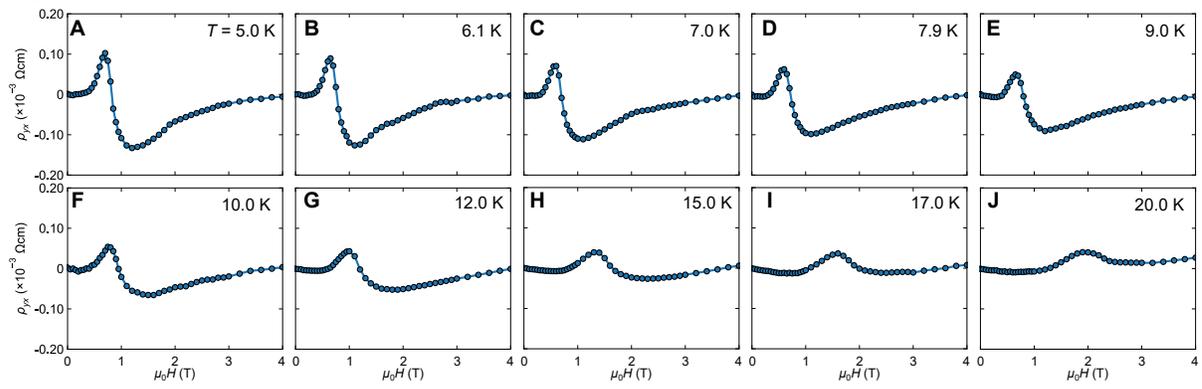

**Magnetic field dependence of the Hall resistivity.** (**A**)-(**J**) Magnetic field dependence of the Hall resistivity at various temperatures up to 20 K measured in a bulk sample.

**Fig. S7.**

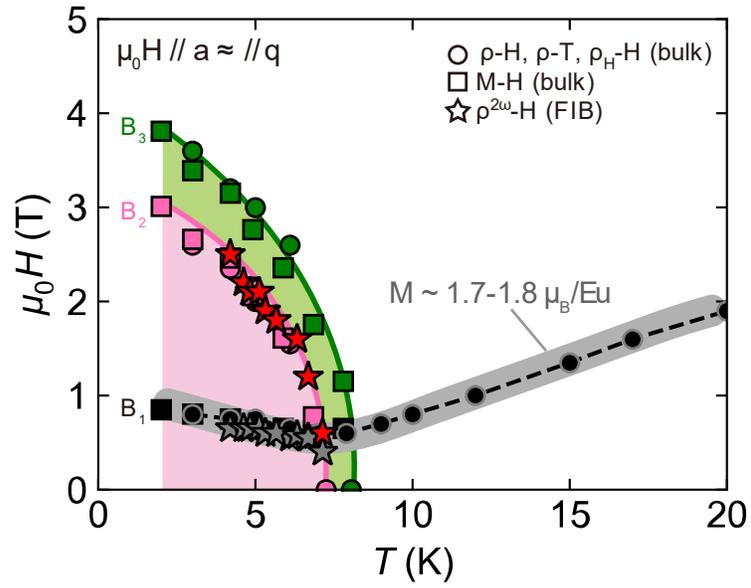

**Magnetic phase diagram and resistivity anomalies.** Field-temperature phase diagram of α-EuP$_3$ for a magnetic field $\mu_0 H$ // $a$, along with the Hall resistivity anomalies shown up to $T$ = 20 K. $B_1$, $B_2$, and $B_3$ denotes the magnetic fields where the anomalies and phase transitions occur as shown in Figure 3.

**Fig. S8.**

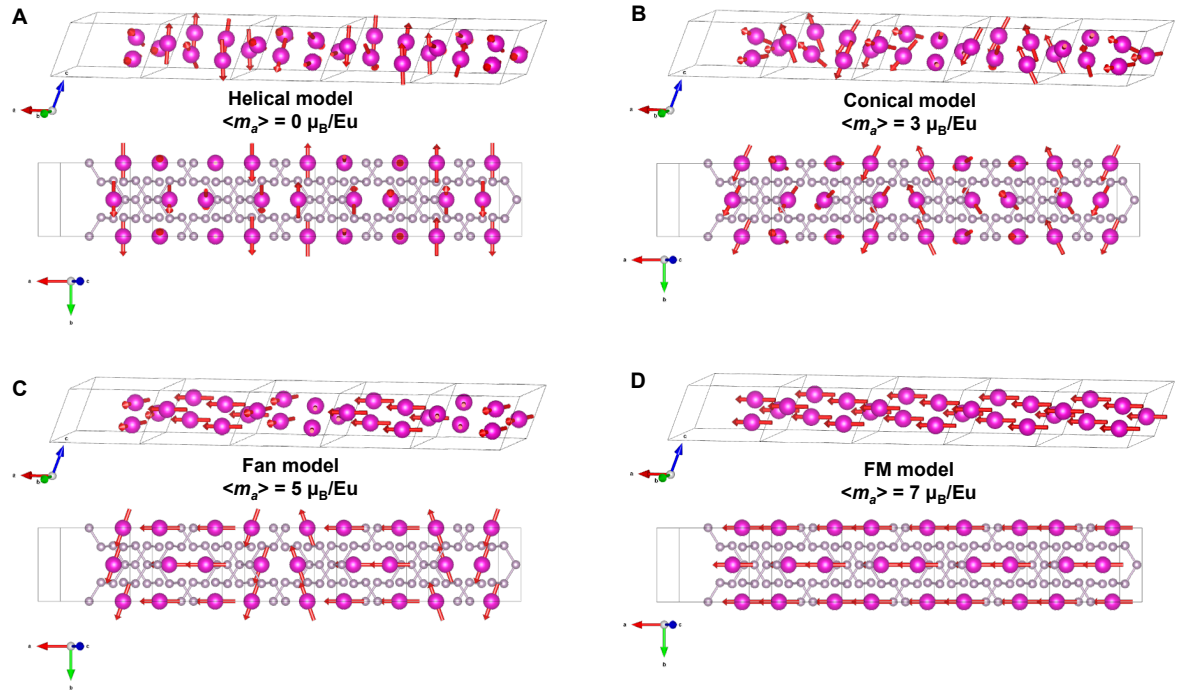

**Simplified magnetic models assuming a propagation vector of** $q$ **= (-0.75, 0, 0)**. Arrangement of the Eu moments in the (**A**) helimagnetic model, (**B**) conical model, (**C**) fan model, and (**D**) ferromagnetic model. Panel **B** shows the case for 3 $\mu_B$/Eu as one example for the conical structure. Only Eu ions are depicted in the top panels for simplicity.

**Table S1.**
**List of the magnetochiral anisotropy coefficient $\gamma$ for various published materials.** TSM and RT stand for topological semimetal and room temperature, respectively. Note that there is a detailed discussion regarding the $\mathcal{P}$-breaking in ZrTe$_5$[S5].

| Material | $|\gamma|$ (m$^2$T$^{-1}$A$^{-1}$) | Crystal structure | Note | Reference |
|---|---|---|---|---|
| ZrTe$_5$ | $4 \times 10^{-7}$ | Polar | TSM, $T = 3$ K | [S5] |
| WTe$_2$ | $3.4 \times 10^{-7}$ | Polar | TSM, $T = 5$ K | [S6] |
| BiTeBr | $3 \times 10^{-12}$ | Polar | $T = 2$ K | [13] |
| Te | $10^{-8}$ | Chiral | RT | [S7] |
| CrNb$_3$S$_6$ | $10^{-12}$ | Chiral | Magnetic, $T = 90$ K | [5] |
| MnSi | $2 \times 10^{-13}$ | Chiral | Magnetic, $T = 35$ K | [4] |
| MnP | $4 \times 10^{-13}$ | Centrosymmetric | Magnetic, $T = 51$ K | [6] |
| MnAu$_2$ | $2 \times 10^{-14}$ | Centrosymmetric | Magnetic, RT | [7] |
| **α-EuP$_3$** | **$6 \times 10^{-11}$** | **Centrosymmetric** | **Magnetic, $T = 4.2$ K** | **This work** |

**SI References**

S1. A. Yanase, A. Hasegawa, Electronic structure of MnP. *J. Phys. C: Solid State Phys.* **13**, 1989 (1980).

S2. J. Xiao, A. Zangwill, M. D. Stiles, Macrospin models of spin transfer dynamics. *Phys. Rev. B* **72**, 014446 (2005).

S3. O. Wessely, B. Skubic, L. Nordström, Spin-transfer torque in helical spin-density waves. *Phys. Rev. B* **79**, 104433 (2009).

S4. V. V. Ustinov, I. A. Yasyulevich, Chirality-dependent spin-transfer torque and current-induced spin rotation in helimagnets. *Phys. Rev. B* **106**, 064417 (2022).

S5. Y. Wang, H. F. Legg, T. Bömerich, J. Park, S. Biesenkamp, A. A. Taskin, M. Braden, A. Rosch, Y. Ando, Gigantic Magnetochiral Anisotropy in the Topological Semimetal $ZrTe_5$. *Phys. Rev. Lett.* **128**, 176602 (2022).

S6. T. Yokouchi, Y. Ikeda, T. Morimoto, Y. Shiomi, Giant Magnetochiral Anisotropy in Weyl Semimetal $WTe_2$ Induced by Diverging Berry Curvature. *Phys. Rev. Lett.* **130**, 136301 (2023).

S7. G. L. J. A. Rikken, N. Avarvari, Strong electrical magnetochiral anisotropy in tellurium. *Phys. Rev. B* **99**, 245153 (2019).